\documentclass[ALICE,manyauthors]{cernphprep}
\usepackage[comma,square,numbers,sort&compress]{natbib}
\usepackage{hyperref}
\usepackage{lineno}
\usepackage{xspace}
\usepackage{color}
\usepackage{siunitx}
\usepackage[utf8]{inputenc} 
\usepackage[T1]{fontenc}
\usepackage{lineno}
\usepackage{epsfig}
\usepackage{color}
\usepackage{xspace}

\usepackage{orcidlink} 

\definecolor{dgreen}{cmyk}{1.,0.,1.,0.1}      
\definecolor{orange}{cmyk}{0.,0.353,1.,0.}    

\newcommand{\com}[1]    {\relax}

\newcommand{\nineH}        {$\sqrt{s}~=~0.9$~Te\kern-.1emV\xspace}
\newcommand{\seven}        {$\sqrt{s}~=~7$~Te\kern-.1emV\xspace}
\newcommand{\twoH}         {$\sqrt{s}~=~0.2$~Te\kern-.1emV\xspace}
\newcommand{\twosevensix}  {$\sqrt{s}~=~2.76$~Te\kern-.1emV\xspace}
\newcommand{\five}         {$\sqrt{s}~=~5.02$~Te\kern-.1emV\xspace}
\newcommand{\twosevensixnn}{$\sqrt{s_{\mathrm{NN}}}~=~2.76$~Te\kern-.1emV\xspace}
\newcommand{\fivenn}       {$\sqrt{s_{\mathrm{NN}}}~=~5.02$~Te\kern-.1emV\xspace}

\newcommand{\MeVc}         {Me\kern-.1emV/$c$\xspace}
\newcommand{\TeV}          {Te\kern-.1emV\xspace}
\newcommand{\GeV}          {Ge\kern-.1emV\xspace}
\newcommand{\MeV}          {Me\kern-.1emV\xspace}

\newcommand{\MeVmass}      {Me\kern-.1emV/$c^2$\xspace}

\def\snn   {\mbox{$\sqrt{s_{_{\rm NN}}}$}}
\newcommand{\pt}{\ensuremath{p_{\rm T}}\xspace}

\graphicspath{{.img/}}

\begin{document}

\begin{titlepage}
\PHyear{2025}       
\PHnumber{080}      
\PHdate{31 March}  

\title{First observation of ultra-long-range azimuthal correlations in low multiplicity pp and p--Pb collisions at the LHC}
\ShortTitle{Ultra-long-range correlations at low multiplicity pp and p--Pb collisions}   

\Collaboration{ALICE Collaboration\thanks{See Appendix~\ref{app:collab} for the list of collaboration members}}
\ShortAuthor{ALICE Collaboration} 

\begin{abstract}

This study presents the first observation of ultra-long-range two-particle azimuthal correlations with pseudorapidity separation of \( |\Delta \eta| > 5.0 \) in proton--proton (pp) and \( |\Delta \eta| > 6.5 \) in proton--lead (p--Pb) collisions at the LHC, down to and below the minimum-bias multiplicity. Two-particle correlation coefficients (${V}_{2\Delta}$) are measured after removing non-flow (jets and resonance decays) contributions using the template-fit method across various multiplicity classes, providing novel insights into the origin of long-range correlations in small systems. Comparisons with the 3D-Glauber + MUSIC + UrQMD hydrodynamic model reveal significant discrepancies at low multiplicities, indicating possible dynamics beyond typical hydrodynamic behavior. Initial-state models based on the Color Glass Condensate framework generate only short-range correlations, while PYTHIA simulations implemented with the string-shoving mechanism also fail to describe these ultra-long-range correlations. The results challenge existing paradigms and question the underlying mechanisms in low-multiplicity pp and p--Pb collisions. The findings impose significant constraints on models describing collective phenomena in small collision systems and advance the understanding of origin of long-range correlations at Large Hadron Collider (LHC) energies.

\end{abstract}
\end{titlepage}

\setcounter{page}{2} 


Prior to the start of the LHC, it was expected that proton--proton collisions at relativistic energies could largely be described as an incoherent sum of parton--parton scatterings. The experimental discovery of significant collective-like effects in high-multiplicity pp and p--Pb collisions at the LHC was, therefore, a big surprise~\cite{Grosse-Oetringhaus:2024bwr, Christiansen:2024bhe}. This led to a large experimental and theoretical program to understand their origin. Collective-like effects are often measured via correlations between particle pairs separated by significant pseudorapidity gaps (\(|\Delta \eta|\)). These long-range correlations (in \(|\Delta \eta|\)) form
structures that are known as ridges~\cite{PHOBOS:2009sau,Aamodt:2011by,ALICE:2022wpn}, and were initially observed in heavy-ion collisions. The ridge, along with two- and multi-particle correlations, anisotropic flow (\( v_2 \)) of identified particles, and strangeness enhancement, has been attributed to the hydrodynamic evolution of the hot and dense quark--gluon plasma (QGP)~\cite{ALICE:2022wpn} formed in heavy-ion collisions. In hydrodynamics, ridge structures are attributed to momentum anisotropy induced by the azimuthally anisotropic geometry of the collision region, which is approximately boost-invariant in rapidity. Interestingly, similar collective-like effects observed in high-multiplicity pp and p–Pb collisions~\cite{CMS:2010ifv,ALICE:2012eyl,Aad:2012gla,ALICE:2014dwt,Aad:2015gqa,CMS:2015yux,Aaij:2015qcq,CMS:2016fnw,ATLAS:2017hap,ALICE:2019zfl,ALICE:2013snk,CMS:2014und,ALICE:2024vzv,ALICE:2016fzo} at LHC energies have sparked debate about whether a hot and dense medium is also formed in these smaller systems~\cite{Grosse-Oetringhaus:2024bwr, Christiansen:2024bhe} or if alternative mechanisms, such as initial-state effects or parton dynamics, are responsible for the observed phenomena. Recent ALICE measurement observed a ridge structure ($|\Delta \eta| < 8.0$) and significant \( v_2 \) over a wide pseudorapidity range ($|\eta| \sim 5.0$) in central p--Pb collisions at \(\sqrt{s_{\mathrm{NN}}} = 5.02~\mathrm{TeV}\)~\cite{ALICE:2023gyf}. The results, supported by hydrodynamic and transport model comparisons, highlight the role of final-state interactions in small systems, similar to those in heavy-ion collisions.

Recent measurements have also revealed long-range correlations in lower multiplicity classes of small collision systems at the LHC~\cite{ALICE:2023ulm}, raising the question of whether a common underlying mechanism drives collective-like effects in high- and low-multiplicity pp as well as p–Pb collisions. While hydrodynamics~\cite{Zhao:2022ayk, ALICE:2024vzv} can describe the ridge and collective behaviors in high-multiplicity pp and p–Pb collisions, its applicability to low-multiplicity events remains uncertain, as these systems may lack sufficient amount of interactions to form a thermalized medium. This raises the critical question of how low the multiplicity can be for hydrodynamics to remain relevant. Additionally, contributions from initial-state effects, which could dominate in the absence of significant final-state interactions in low-multiplicity collisions, need further exploration. Recent ALICE measurements of ridge yields in pp collisions~\cite{ALICE:2023ulm} at event multiplicities down to the average multiplicity of minimum-bias triggered events (minimum-bias multiplicity) show significantly higher values compared to \(e^+e^-\)  collisions at similar multiplicities, where initial-state momentum and geometry anisotropies are absent. This suggests that initial-state momentum anisotropy may contribute to the ridge in small collision systems. Previously, the Color Glass Condensate (CGC) effective theory predicted that long-range correlations in small systems might arise from initial state momentum anisotropies in the colliding ions~\cite{Dusling:2017dqg}. However, recent calculations within the 3+1D IP-Glasma framework suggest that initial-state momentum correlations are relatively short-range in pseudorapidity ($|\Delta \eta| \lesssim 3.0$)~\cite{Schenke:2022mjv}, whereas the initial geometry can generate ultra-long-range correlations extending beyond $|\Delta \eta| > 5.0$. In ALICE, previous correlation measurements extending down to low-multiplicity pp collisions were performed at midrapidity and were limited to $|\Delta \eta| < 1.8$ due to the acceptance of the tracking detectors~\cite{ALICE:2023ulm}. ATLAS and CMS have also observed finite near-side long-range correlations in the range of $2.0 < |\Delta \eta| < 5.0$~\cite{Aad:2015gqa, CMS:2015fgy}. In these ranges, both initial geometry-driven hydrodynamics evolution and initial momentum correlations could contribute to these observations. Studying ultra-long-range correlations ($|\Delta \eta| > 5.0$ or 6.5) in pp and p–Pb collisions, down to and below minimum-bias multiplicity, could offer critical insights into the contributions of initial- and final-state effects, possibly helping to disentangle the different scenarios at play.

Notably, the general features of pp collisions, such as jets and multiplicity distributions, have traditionally been described by dynamical models based on string or cluster hadronization, as implemented in PYTHIA8~\cite{Bierlich:2022pfr}. This jet- and underlying-event-dominated paradigm differs fundamentally from the initial geometry-driven hydrodynamic or CGC approaches, which are used to explain long-range correlations in high-multiplicity events. Any possible observation of ultra-long-range correlations down to low multiplicities would challenge this paradigm and raise important questions about the underlying physics at the low multiplicity range. Interestingly, in the string-shoving version of PYTHIA8~\cite{Bierlich:2017vhg, Bierlich:2014xba}, multiple partonic subcollisions create a dense system of strings. The overlapping strings generate a transverse pressure that mimics transverse flow, generating the ridge structure in high-multiplicity pp collisions without any QGP formation.  The strength of this effect in pseudorapidity space is directly linked to the spatial extension of the strings. Testing these alternative models at low multiplicities using ultra-long-range correlations could help further constrain these models and improve our understanding of the origin of collective phenomena across different multiplicity classes in small systems.

In this Letter, the first measurements of ultra-long-range two-particle correlations (\(|\Delta \eta| > 5.0\) and \(|\Delta \eta| > 6.5\)) in pp and p--Pb collisions at the LHC, reaching down to or even below minimum-bias multiplicities, are presented. These measurements are compared with 3+1D hydrodynamic model~\cite{Zhao:2022ayk} and PYTHIA string-shoving model estimations to explore their possible origins.

The analyzed data samples consist of pp collisions at $\sqrt{s}=13$~TeV and p--Pb collisions at $\snn=5.02$~TeV, collected by the ALICE detector during the LHC Run 2 campaign (2015--2018). The V0 detector~\cite{ALICE:2013axi}, comprising two scintillator arrays at pseudorapidity ranges $2.8 < \eta < 5.1$ and $-3.7 < \eta < -1.7$, was used for triggering and event selection. Minimum-bias (MB) events were selected by requiring signals from at least one charged particle in each V0 counter~\cite{ALICE:2008ngc, ALICE:Perform}, while the high-multiplicity (HM) data sample consists of the top 0.07\% of events with the highest V0 signal, selected using a dedicated HM trigger. Multiplicity classes are defined based on
the number of charged tracks ($N_{\rm ch}$) detected within the Time Projection
Chamber (TPC)~\cite{ALICE:2008ngc,ALICE:Perform}, considering tracks with $|\eta|<0.8$
and transverse momentum in the range $0.2 < \pt < 3$~GeV/$c$. For pp collisions, the HM trigger selects events which typically have $N_{\rm ch} > 40$. The p--Pb collision sample consists only of MB events. For both pp and p--Pb collisions, events were required to have a reconstructed primary vertex within \unit {10} {cm} of the nominal interaction point. The selected dataset includes $\sim8.7 \times 10^7$ HM and $\sim5.2 \times 10^8$ MB pp collisions, with integrated luminosities of $\sim$1.5~$\text{nb}^{-1}$ and $\sim$9~$\text{nb}^{-1}$, respectively~\cite{ALICE-PUBLIC-2021-005}. For p--Pb collisions, the dataset comprises $\sim5 \times 10^8$ MB collisions with an integrated luminosity of 0.24~$\text{nb}^{-1}$~\cite{ALICE:2014gvw}.

The observable used to characterize ultra-long-range correlations is derived from two-particle correlations (2PC)~\cite{ALICE:2013snk}, normalized by the number of trigger particles. This per-trigger yield is constructed as a function of the azimuthal angle difference, \( \Delta\varphi \), and the pseudorapidity difference, \( \Delta\eta \), between trigger and associated particles and is measured as
\begin{equation}
\label{eq:twocorr}
\frac{1}{N_{\mathrm{trig}}}\frac{\mathrm{d}^{2}N_{\mathrm{pair}}}{\mathrm{d}\Delta\eta\mathrm{d}\Delta\varphi} = \frac{S(\Delta\eta,\Delta\varphi)}{B(\Delta\eta,\Delta\varphi)},
\end{equation}
where $S(\Delta\eta,\Delta\varphi)$ and $B(\Delta\eta,\Delta\varphi)$ are the same and mixed event distributions constructed using particles selected from the Forward Multiplicity Detector (FMD), which covers forward $1.7 < \eta < 5.1$ (FMD1,2) and backward $-3.1 < \eta < -1.7$ (FMD3) rapidity regions~\cite{ALICE:2008ngc,ALICE:Perform}. The pair-acceptance effect due to the finite detector size and inefficiencies is corrected by dividing \( S(\Delta\eta, \Delta\varphi) \) by an adequately normalized \( B(\Delta\eta, \Delta\varphi) \)~\cite{ALICE:2012eyl}. The $S(\Delta\eta,\Delta\varphi)$ is also normalized with the number of trigger particles. Since the FMD is not a tracking detector, it counts all particles within its acceptance irrespective of their \( p_{\mathrm{T}} \). In p--Pb collisions, particles from two different combinations of the FMD sections are used: (a) $2.6 < \eta < 3.2$ (FMD1,2), $-3.0 < \eta < -2.0$ (FMD3), and (b) $3.8 < \eta < 4.8$ (FMD1,2), $-3.0 < \eta < -2.0$ (FMD3). In pp collisions, part of FMD1,2 was inactive, liming the acceptance range
to  $2.6 < \eta < 3.2$ (FMD1,2) and $-3.0 < \eta < -2.0$ (FMD3). These selected FMD intervals allow $\Delta\eta$ separations between particle pairs of $5.0 < |\Delta\eta| < 6.0$ in pp collisions, and $5.0 < |\Delta\eta| < 6.0$ as well as $6.5 < |\Delta\eta| < 7.5$ in p--Pb collisions.

The per-trigger yield contains both flow and non-flow effects (from jets and resonance decays). To minimize non-flow effects, the template-fit method~\cite{Aad:2015gqa} is applied to the $\Delta\varphi$ projections of the 2D per-trigger yield. The $\Delta\varphi$ projections are calculated as
\begin{equation}
\label{eq:pertrigg}
Y(\Delta\varphi) = \frac{1}{N_{\mathrm{trig}}}\frac{\mathrm{d}N_{\mathrm{pair}}}{\mathrm{d}\Delta\varphi} = \int_{|\Delta\eta|>5.0/6.5}\left(\frac{1}{{N}_{\mathrm{trig}}}\frac{\mathrm{d}^{2}N_{\mathrm{pair}}}{\mathrm{d}\Delta\eta\mathrm{d}\Delta\varphi}\right)  \mathrm{d}\Delta\eta.
\end{equation}

The low-multiplicity (LM) events are used as a template~\cite{Aad:2015gqa} to fit higher multiplicity classes, assuming higher multiplicity events are expressed as superposition of LM events with additional flow contributions as shown in Eq.~(\ref{eq:eqA1}).
\begin{equation}
    Y(\Delta\varphi) = FY^{\rm LM}(\Delta\varphi)+G \left[ 1+\sum_{n=2}^{3}2{V}_{n\Delta} \cos (n\Delta\varphi)  \right].
    \label{eq:eqA1}
\end{equation}
Here, $Y(\Delta\varphi)$ and $Y^{\rm LM}(\Delta\varphi)$ are the one-dimensional $\Delta\varphi$ projections in the HM and LM event classes, with $F$ and $G$ being the scaling factors, and the additional flow contribution in the HM event classes described by the Fourier distribution. The $|\Delta\eta|$ ranges used for the $\Delta\varphi$ projections are $|\Delta\eta| > 5.0$ for pp and both $|\Delta\eta| > 5.0$ and $|\Delta\eta| > 6.5$  for p--Pb collisions. The two-particle correlation coefficients, ${V}_{n\Delta}$, are estimated by fitting the $\Delta\varphi$ projections of the HM event classes using Eq.~(\ref{eq:eqA1}).

The ultra-long-range correlations of $|\Delta\eta| > 5.0$ in pp collisions and $|\Delta\eta| > 6.5$ in p--Pb collisions are not affected by non-flow contributions from relatively short-range resonance decays and jet fragmentation on the near side ($\Delta\varphi < \pi/2$) of the per-trigger yield. The jet contributions on the away side ($\Delta\varphi > \pi/2$) have been taken into account using the non-flow subtraction with the template fit method to extract ${V}_{2\Delta}$. The definition of the ${V}_{2\Delta}$ coefficients in this work differs from the previous ALICE measurement of near-side jet yields~\cite{ALICE:2023ulm}, which was restricted to $1.4 < |\Delta\eta| < 1.8$ in the TPC at midrapidity ($-0.9 < \eta < 0.9$) without non-flow removal. The \( V_{2\Delta} \) observable presented in this work also differs from \( v_2 \) in Ref.~\cite{ALICE:2023gyf}, which was estimated using three pairs of two-particle correlations (3x2PC method)~\cite{PHENIX:2021ubk}. In that approach, the \( v_2 \) was obtained by multiplying \( V_{2\Delta}^{\rm TPC-FMD1,2} \) and \( V_{2\Delta}^{\rm TPC-FMD3} \), and then dividing by \( V_{2\Delta}^{\rm FMD1,2-FMD3} \), assuming flow factorization~\cite{Zhao:2022ayk, ALICE:2023gyf}. This procedure partially cancels out the effects of longitudinal flow decorrelation. This is inconvenient for data--model comparisons, as theoretical models may significantly overestimate or underestimate individual \( V_{2\Delta} \) measurements. However, the cancellation effects inherent in the 3x2PC method make \( v_2 \) less sensitive to flow decorrelations, potentially leading to accidental agreements or discrepancies between data and models. In contrast, the \( V_{2\Delta} \) observable is more sensitive to longitudinal flow decorrelation effects and does not rely on any assumptions related to flow factorization, thereby providing tighter constraints on theoretical models.

Systematic uncertainties in the \( V_{2\Delta} \) measurement are evaluated by varying the event and track selection criteria from their default settings.
The difference between the
default results and the ones from the variations is calculated across \( N_{\rm ch} \) intervals using the Barlow criterion~\cite{Barlow:2002yb}. Variations with Barlow differences exceeding 1$\sigma$ for more than one-third of \( N_{\rm ch} \) intervals are included in the uncertainty for all \( N_{\rm ch} \) intervals. Varying the selected range of primary vertex position along the beam axis from \( |z_{\text{vtx}}| < 10 \, \text{cm} \) to \( 8 \, \text{cm} \) introduces uncertainties of \(\sim 2\%\) for p--Pb and \(\sim 1\%\) for pp collisions. Variations in the correlation between V0 and FMD multiplicities, used to reject pileup, contribute with a \(\sim 1\%\) uncertainty for p--Pb and up to \(\sim 7\%\) for pp. Changes in track selection for \( N_{\text{ch}} \) estimation add \(\sim 3\%\) uncertainty in both systems. The FMD multiplicity is significantly affected by secondary particles resulting from the scattering of primary particles on detectors and materials present along the path to the FMD detector~\cite{ALICE:2004ftm}. Since the FMD lacks tracking capabilities, primary particles cannot be distinguished from secondary ones. Correction factors, estimated using AMPT and EPOS-LHC event generators~\cite{Lin:2004en, Pierog:2013ria}, are defined as the ratio of \( V_{2\Delta} \) for primary particles to that for all reconstructed particles, including secondaries. The AMPT-based correction factors (up to \(\sim 30\%\)) are applied to the reconstructed EPOS-LHC simulation and compared to those at the generated level, with the difference assigned as the residual non-closure (\(\sim 3\%\)) in the measurement. The effect of varying the material budget (such as support structures, shielding, etc.) is assessed using AMPT-generated particles passed through a GEANT3~\cite{Brun:1994aa} simulation of the ALICE detector and contributes approximately \(\sim 3\%\). Non-flow effects are assessed by varying the \( N_{\text{ch}} \) width of the  LM template (\( 0\text{--}5 \) to \( 0\text{--}4 \) in p--Pb,  \( 0\text{--}10 \) to \( 0\text{--}5 \) in pp), leading to \(\sim 2\%\) uncertainty in p--Pb and \(\sim 7\%\) in pp collisions. Residual non-flow, evaluated with template fits to PYTHIA8-generated correlations, ranges from \(\sim 10\%\) at high multiplicity to \(\sim 17\%\) at low multiplicity. Differences between the results obtained using the 2017 and 2018 pp datasets lead to a \(\sim 12\%\) uncertainty in the measured \( V_{2\Delta} \) values. All contributions are added in quadrature to obtain the total systematic uncertainty.

\begin{figure}[tb]
    \begin{center}
    \includegraphics[width = 0.62\textwidth]{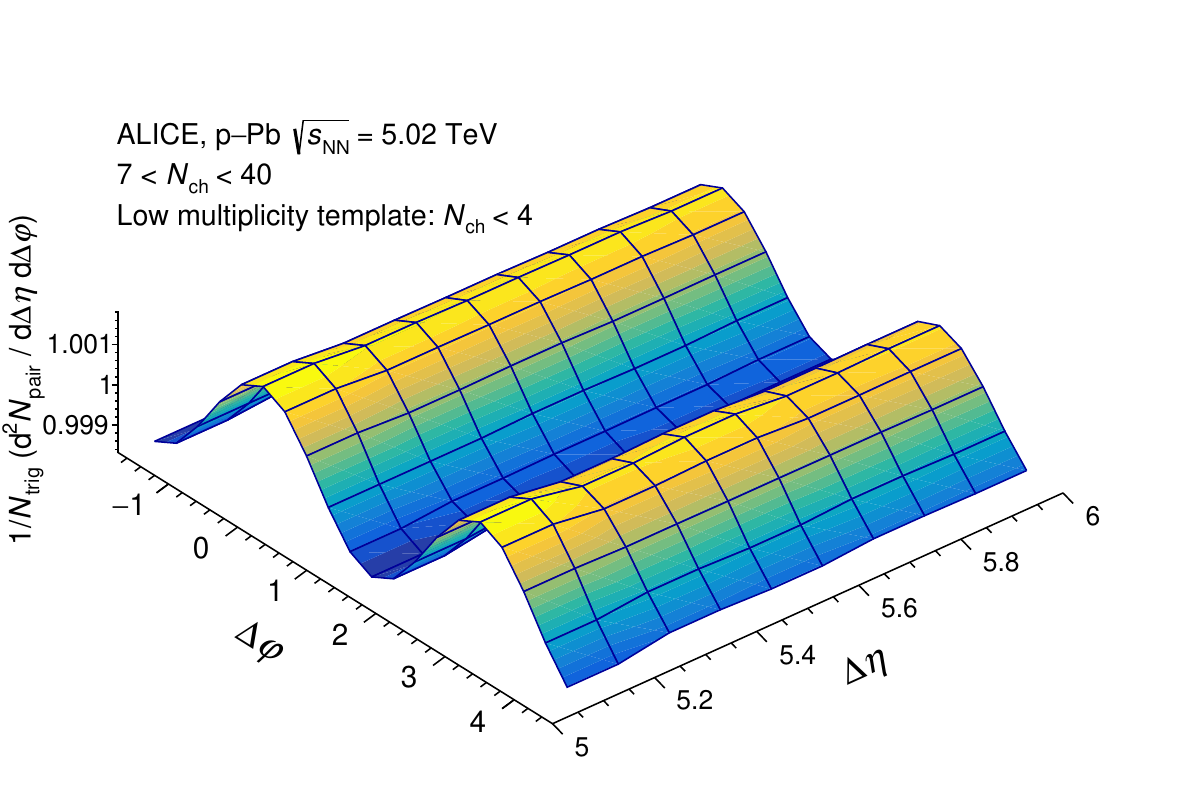}
   \end{center}
    \caption{The ultra-long-range (\(|\Delta \eta| > 5.0\)) per-trigger yield, measured as a function of \(\Delta \eta\) and \(\Delta \varphi\) in p--Pb collisions at \(\sqrt{s_{\text{NN}}} = 5.02\)~TeV for \(7 < N_{\text{ch}} < 40\), shows a double-ridge structure after non-flow removal using the template-fit method.}    
\label{fig:2Dridge}
\end{figure}

\begin{figure}[tbh]
    \begin{center}
    \includegraphics[width = 0.54\textwidth]{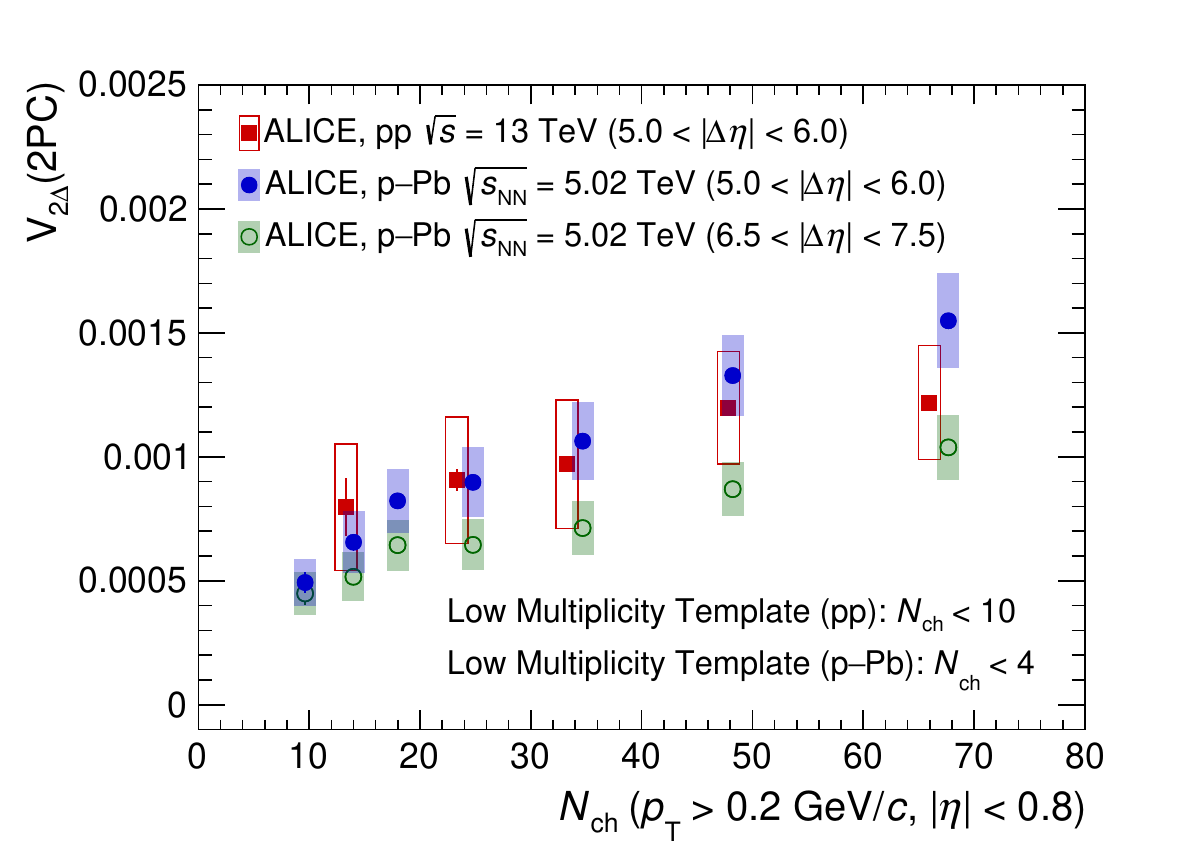}
    \end{center}
    \caption{The \(N_{\text{ch}}\) dependence of the second-order two-particle correlation coefficient \( V_{2\Delta} (\text{2PC}) \), estimated using the template-fit method, in pp (\(|\Delta \eta| > 5.0\)) collisions at \(\sqrt{s} = 13\) TeV and p--Pb (\(|\Delta \eta| > 5.0\) and \(|\Delta \eta| > 6.5\)) collisions at \(\sqrt{s_{\text{NN}}} = 5.02\) TeV.
}
    \label{fig:alldatapoints}
\end{figure}

The ultra-long-range double-ridge structure in p--Pb collisions for \( 7 < N_{\text{ch}} < 40 \), after non-flow subtraction using the template-fit method, is shown in Fig.~\ref{fig:2Dridge}. This is the first observation of a double-ridge correlation extending to \( |\Delta \eta| > 5.0 \) at \( \langle N_{\text{ch}} \rangle \approx 21 \), close to the minimum-bias multiplicity (\( \langle N_{\text{ch}} \rangle \approx 24 \)) for the selected p--Pb collisions in this analysis. This demonstrates that the double ridge in p--Pb collisions, previously observed in the higher multiplicity classes~\cite{ALICE:2013snk, ALICE:2023gyf}, also exists down to the minimum-bias multiplicity after non-flow removal.  Figure~\ref{fig:alldatapoints} presents the \( V_{2\Delta} \) values obtained using the template-fit method for pp (\( 5.0 < |\Delta \eta| < 6.0 \)), p--Pb (\( 5.0 < |\Delta \eta| < 6.0 \)), and p--Pb (\( 6.5 < |\Delta \eta| < 7.5 \)) collisions as a function of \( N_{\text{ch}} \). In all cases, \( V_{2\Delta} \) increases with \( N_{\text{ch}} \), showing a higher correlation strength at higher multiplicities. For pp collisions, the lowest \( N_{\text{ch}} \) bin is close to the minimum-bias multiplicity (\( \langle N_{\text{ch}} \rangle \approx 10 \)) and \( V_{2\Delta} \) is non-zero with a significance of 3.1 standard deviation (\(\sigma \)). For p--Pb collisions, the lowest \( N_{\text{ch}} \) (7--12) bin corresponds to a multiplicity lower than the minimum-bias value (\( \langle N_{\text{ch}} \rangle \approx 24 \)) and the \( V_{2\Delta} \) values are non-zero with a significance of \( 5.2\sigma \) for both \( |\Delta \eta| > 5.0 \) and \( |\Delta \eta| > 6.5 \) cases. In p--Pb collisions, \( V_{2\Delta} \)(\( 5.0 <|\Delta \eta| <6.0 \)) is consistent with \( V_{2\Delta} \)(\( 6.5 <|\Delta \eta| <7.5 \)) up to \( N_{\text{ch}} \approx 20 \) within measurement uncertainties, as shown in  Fig.~\ref{fig:alldatapoints}. At higher \( N_{\text{ch}} \), \( V_{2\Delta} \) measured for \( 5.0 < |\Delta \eta| < 6.0 \) is systematically larger than that for \( 6.5 < |\Delta \eta| < 7.5 \), with the difference amounting to \(\sim 2.2\sigma\) for the results at \( N_{\text{ch}} > 40 \). This indicates the possible sensitivity of \( V_{2\Delta} \) to non-trivial longitudinal dynamics and its evolution with multiplicity and \( |\Delta \eta| \) separation between the correlated pairs in small collision systems.

\begin{figure}[tbh]
    \includegraphics[width = 0.496\textwidth]{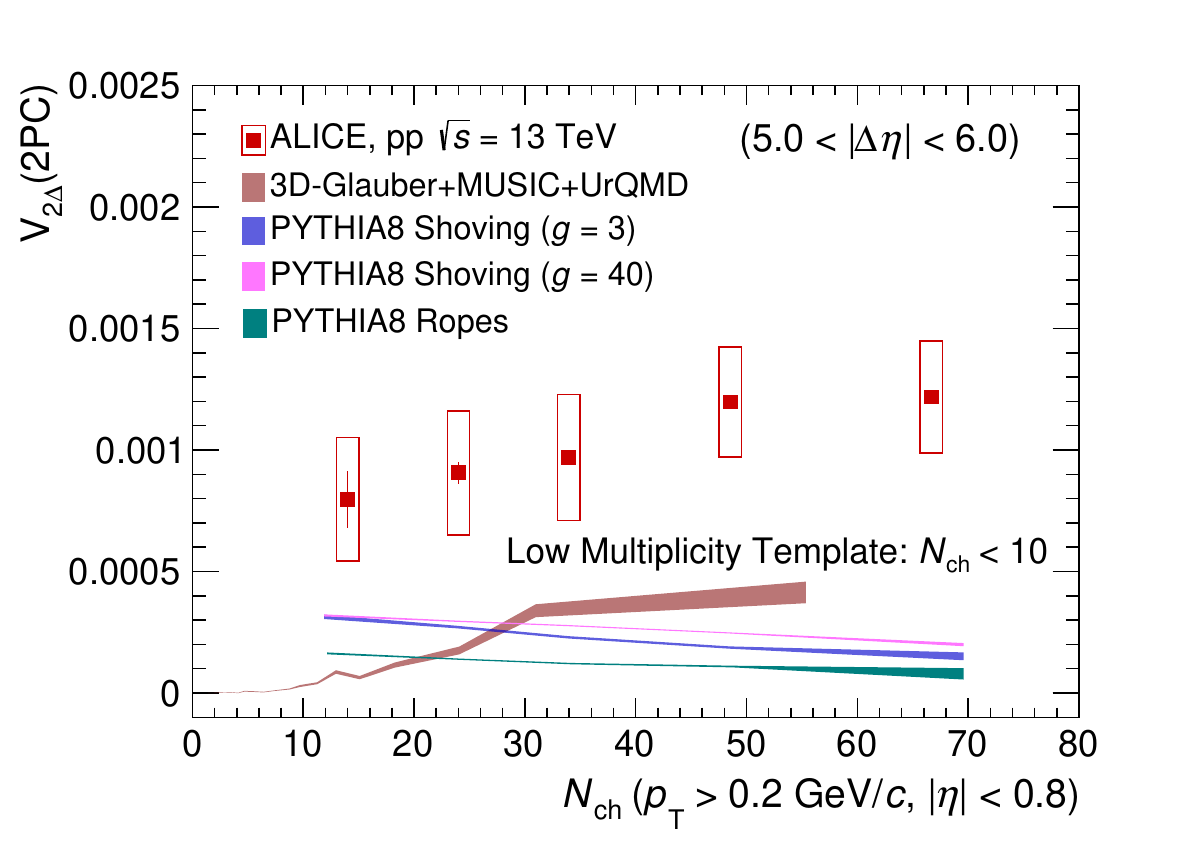}
    \includegraphics[width = 0.5\textwidth]{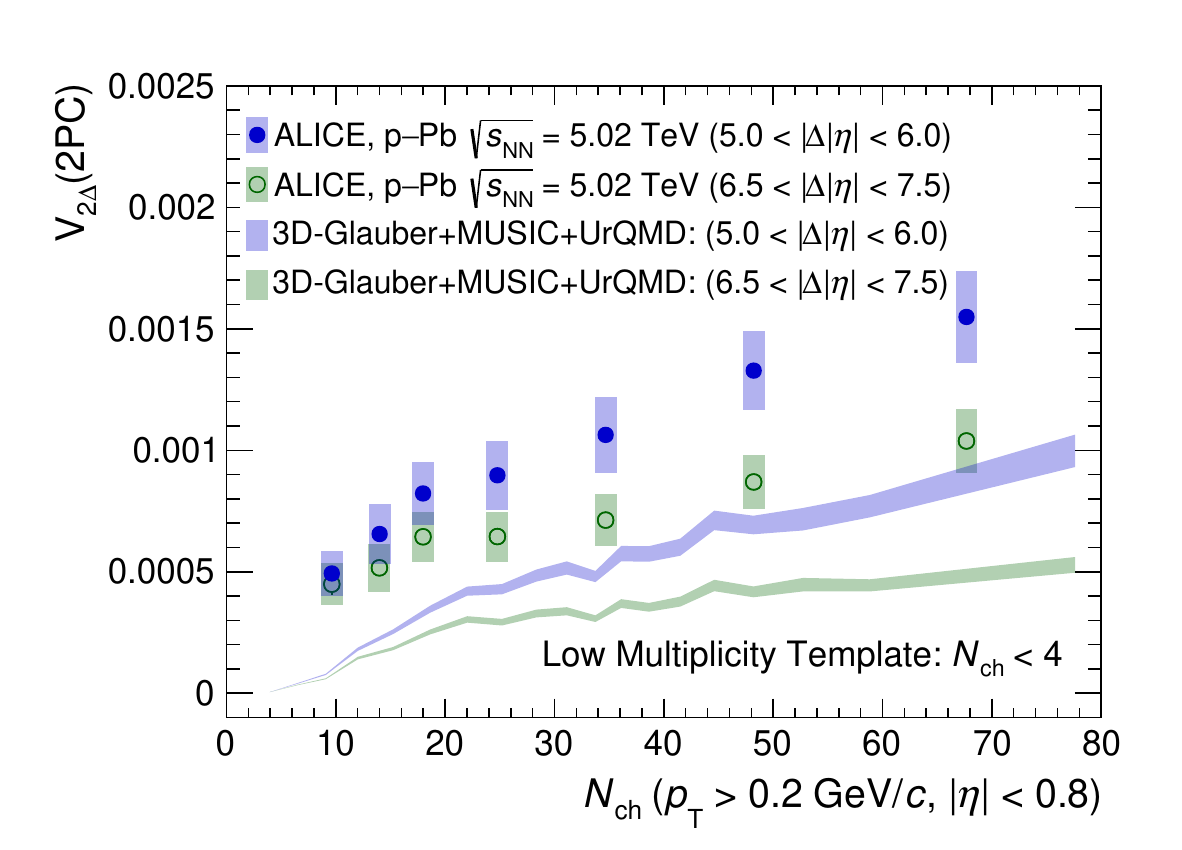}
    \caption{Left: The \( N_{\rm ch} \) dependence of the second-order two-particle correlation coefficient \( V_{2\Delta} (\rm{2PC}) \), estimated using the template-fit method, in pp (\( |\Delta \eta| > 5.0 \)) collisions at \( \sqrt{s} = 13 \)~TeV, and compared with predictions from the 3D-Glauber + MUSIC + UrQMD model~\cite{Zhao:2022ugy}, as well as with PYTHIA8 variations with string-shoving~\cite{Bierlich:2017vhg, Bierlich:2014xba} and ropes~\cite{Bierlich:2022ned} implementations. Right: The $N_{\rm ch}$ dependence of the \( V_{2\Delta} (\rm{2PC}) \), estimated using template-fit method, in p--Pb ($|\Delta \eta| > 5.0$ and $|\Delta \eta| > 6.5$) collisions at $\snn=5.02$~TeV and its comparison with the 3D-Glauber + MUSIC + UrQMD predictions.}
    \label{fig:datamodelcomp}
\end{figure}

To interpret the results shown in Fig.~\ref{fig:alldatapoints}, a comparison is made with the 3D-Glauber + MUSIC + UrQMD model~\cite{Zhao:2022ugy}, as shown in Fig.~\ref{fig:datamodelcomp}. This 3+1D hydrodynamic framework combines 3D-Glauber initial conditions, viscous hydrodynamics (MUSIC), and the UrQMD model for hadronic interactions. In this model, the elliptic flow (\( v_2 \)) arises from the 3D hydrodynamic evolution driven by the anisotropy of the initial geometry. The framework describes reasonably well measurements such as particle production (\( \rm{d}\it{N}_{\text{ch}}/\rm{d}\eta \)), multiplicity dependence of the average transverse momentum $(\langle p_{\text{T}} \rangle)$ of identified particles, \( N_{\text{ch}} \)- and \( p_{\text{T}} \)-dependence of \( v_2 \) estimated from relatively short-range correlations (\(|\Delta \eta| <5.0 \)) as well as flow fluctuations/decorrelations in p--Pb collisions at 5.02 TeV~\cite{Zhao:2022ayk, Shen:2022oyg, ALICE:2023gyf}. However, the comparison shown in Fig.~\ref{fig:datamodelcomp} reveals that while the \( V_{2\Delta} \) in the model increases with \( N_{\text{ch}} \), the experimental \( V_{2\Delta} \) values consistently exceed the model expectations across all multiplicity classes in both pp and p--Pb collisions. 
Notably, in p--Pb collisions, the \( V_{2\Delta} \) values are approximately 3\text{--}4 times larger than the hydrodynamic predictions in the lower multiplicity classes, with the discrepancy reducing to a factor of about 1.5\text{--}2.0 in the higher multiplicity classes. Assuming the applicability of hydrodynamics across multiplicities in small collision systems, initial-state effects --- including rapidity decorrelations in the collision geometry --- play a key role in the development of long-range correlations~\cite{Zhao:2022ayk}. However, these effects are still not well understood in small collision systems.

The data-model comparison shown in  Fig.~\ref{fig:datamodelcomp} indicates that in very low-multiplicity pp and p–Pb collisions, the \( V_{2\Delta} \) estimated from ultra-long-range correlations may not arise from the hydrodynamic expansion of the system, or at least not with the initial conditions implemented in the 3D-Glauber + MUSIC + UrQMD model shown here. Also, the CGC framework, which generates short-range azimuthal correlations with the effects vanishing at ultra-long ranges~\cite{Schenke:2022mjv}, fails to describe the measurements presented in this paper.

In addition, the \( N_{\text{ch}} \) dependence of \( V_{2\Delta} \) (\( 5.0 < |\Delta \eta| < 6.0 \)) in pp collisions is compared with estimations from the string-shoving implementation in PYTHIA8~\cite{Bierlich:2017vhg, Bierlich:2014xba} in  Fig.~\ref{fig:datamodelcomp}. Two values of the shoving parameter (\( \it{g}\) = 3  and \( \it{g}\) = 40) are used, with a higher $\it{g}$ value corresponding to a stronger repulsive interaction between overlapping color strings. The results show that string shoving in PYTHIA8 produces a decreasing \( V_{2\Delta} \) with increasing \( N_{\text{ch}} \), which is qualitatively opposite to the trend observed in the data. Additionally, while the PYTHIA8 model with color ropes enabled in the hadronization process~\cite{Bierlich:2022ned} successfully describes strangeness enhancement in pp collisions, it fails to generate significant long-range correlations, as shown in  Fig.~\ref{fig:datamodelcomp}. These discrepancies suggest that the string-shoving or ropes mechanism alone are insufficient to describe the ultra-long-range correlations observed in pp collisions, even at higher multiplicities, highlighting the need for additional mechanisms or theoretical refinements to fully explain ultra-long-range correlations in small systems.

This paper presents the first observation of ultra-long-range correlations in $5.0 < |\Delta \eta| < 6.0$ (in pp and p--Pb collisions) and $|\Delta \eta| > 6.5$ (in p--Pb collisions) down to and below minimum-bias multiplicities at the LHC. These results challenge two paradigms at once: the extent to which hydrodynamics, successful in explaining many collective-like effects in high-multiplicity pp and p–Pb collisions, applies as multiplicity decreases, and whether PYTHIA remains an effective approach for jet-dominated, low-multiplicity small collision systems. The use of ultra-long-range correlations, combined with template fitting for
non-flow removal, provides a uniquely robust measurement that is minimally affected by non-flow and CGC effects. It remains highly sensitive to non-trivial longitudinal dynamics and offers strong resolving power between different possible physical mechanisms (e.g., hydrodynamics, CGC, PYTHIA8 Shoving, etc.) contributing to long-range correlations in small systems. Data–model comparisons presented in this paper reveal that the mechanisms considered so far are unable to explain the origin of these ultra-long-range correlations in pp and p–Pb collisions at the LHC. Therefore, this work presents unprecedented constraints on models aiming to explain collective-like effects in small collision systems, from high to low multiplicity, in a consistent way.


\newenvironment{acknowledgement}{\relax}{\relax}
\begin{acknowledgement}
\section*{Acknowledgements}

The ALICE Collaboration would like to thank all its engineers and technicians for their invaluable contributions to the construction of the experiment and the CERN accelerator teams for the outstanding performance of the LHC complex.
The ALICE Collaboration gratefully acknowledges the resources and support provided by all Grid centres and the Worldwide LHC Computing Grid (WLCG) collaboration.
The ALICE Collaboration acknowledges the following funding agencies for their support in building and running the ALICE detector:
A. I. Alikhanyan National Science Laboratory (Yerevan Physics Institute) Foundation (ANSL), State Committee of Science and World Federation of Scientists (WFS), Armenia;
Austrian Academy of Sciences, Austrian Science Fund (FWF): [M 2467-N36] and Nationalstiftung f\"{u}r Forschung, Technologie und Entwicklung, Austria;
Ministry of Communications and High Technologies, National Nuclear Research Center, Azerbaijan;
Conselho Nacional de Desenvolvimento Cient\'{\i}fico e Tecnol\'{o}gico (CNPq), Financiadora de Estudos e Projetos (Finep), Funda\c{c}\~{a}o de Amparo \`{a} Pesquisa do Estado de S\~{a}o Paulo (FAPESP) and Universidade Federal do Rio Grande do Sul (UFRGS), Brazil;
Bulgarian Ministry of Education and Science, within the National Roadmap for Research Infrastructures 2020-2027 (object CERN), Bulgaria;
Ministry of Education of China (MOEC) , Ministry of Science \& Technology of China (MSTC) and National Natural Science Foundation of China (NSFC), China;
Ministry of Science and Education and Croatian Science Foundation, Croatia;
Centro de Aplicaciones Tecnol\'{o}gicas y Desarrollo Nuclear (CEADEN), Cubaenerg\'{\i}a, Cuba;
Ministry of Education, Youth and Sports of the Czech Republic, Czech Republic;
The Danish Council for Independent Research | Natural Sciences, the VILLUM FONDEN and Danish National Research Foundation (DNRF), Denmark;
Helsinki Institute of Physics (HIP), Finland;
Commissariat \`{a} l'Energie Atomique (CEA) and Institut National de Physique Nucl\'{e}aire et de Physique des Particules (IN2P3) and Centre National de la Recherche Scientifique (CNRS), France;
Bundesministerium f\"{u}r Bildung und Forschung (BMBF) and GSI Helmholtzzentrum f\"{u}r Schwerionenforschung GmbH, Germany;
General Secretariat for Research and Technology, Ministry of Education, Research and Religions, Greece;
National Research, Development and Innovation Office, Hungary;
Department of Atomic Energy Government of India (DAE), Department of Science and Technology, Government of India (DST), University Grants Commission, Government of India (UGC) and Council of Scientific and Industrial Research (CSIR), India;
National Research and Innovation Agency - BRIN, Indonesia;
Istituto Nazionale di Fisica Nucleare (INFN), Italy;
Japanese Ministry of Education, Culture, Sports, Science and Technology (MEXT) and Japan Society for the Promotion of Science (JSPS) KAKENHI, Japan;
Consejo Nacional de Ciencia (CONACYT) y Tecnolog\'{i}a, through Fondo de Cooperaci\'{o}n Internacional en Ciencia y Tecnolog\'{i}a (FONCICYT) and Direcci\'{o}n General de Asuntos del Personal Academico (DGAPA), Mexico;
Nederlandse Organisatie voor Wetenschappelijk Onderzoek (NWO), Netherlands;
The Research Council of Norway, Norway;
Pontificia Universidad Cat\'{o}lica del Per\'{u}, Peru;
Ministry of Science and Higher Education, National Science Centre and WUT ID-UB, Poland;
Korea Institute of Science and Technology Information and National Research Foundation of Korea (NRF), Republic of Korea;
Ministry of Education and Scientific Research, Institute of Atomic Physics, Ministry of Research and Innovation and Institute of Atomic Physics and Universitatea Nationala de Stiinta si Tehnologie Politehnica Bucuresti, Romania;
Ministerstvo skolstva, vyskumu, vyvoja a mladeze SR, Slovakia;
National Research Foundation of South Africa, South Africa;
Swedish Research Council (VR) and Knut \& Alice Wallenberg Foundation (KAW), Sweden;
European Organization for Nuclear Research, Switzerland;
Suranaree University of Technology (SUT), National Science and Technology Development Agency (NSTDA) and National Science, Research and Innovation Fund (NSRF via PMU-B B05F650021), Thailand;
Turkish Energy, Nuclear and Mineral Research Agency (TENMAK), Turkey;
National Academy of  Sciences of Ukraine, Ukraine;
Science and Technology Facilities Council (STFC), United Kingdom;
National Science Foundation of the United States of America (NSF) and United States Department of Energy, Office of Nuclear Physics (DOE NP), United States of America.
In addition, individual groups or members have received support from:
Czech Science Foundation (grant no. 23-07499S), Czech Republic;
FORTE project, reg.\ no.\ CZ.02.01.01/00/22\_008/0004632, Czech Republic, co-funded by the European Union, Czech Republic;
European Research Council (grant no. 950692), European Union;
Deutsche Forschungs Gemeinschaft (DFG, German Research Foundation) ``Neutrinos and Dark Matter in Astro- and Particle Physics'' (grant no. SFB 1258), Germany;
ICSC - National Research Center for High Performance Computing, Big Data and Quantum Computing and FAIR - Future Artificial Intelligence Research, funded by the NextGenerationEU program (Italy).

\end{acknowledgement}

\bibliographystyle{utphys}   
\bibliography{bibliography}

\newpage
\appendix

%
%

\section{The ALICE Collaboration}
\label{app:collab}
\begin{flushleft} 
\small

S.~Acharya\,\orcidlink{0000-0002-9213-5329}\,$^{\rm 50}$, 
A.~Agarwal$^{\rm 133}$, 
G.~Aglieri Rinella\,\orcidlink{0000-0002-9611-3696}\,$^{\rm 32}$, 
L.~Aglietta\,\orcidlink{0009-0003-0763-6802}\,$^{\rm 24}$, 
M.~Agnello\,\orcidlink{0000-0002-0760-5075}\,$^{\rm 29}$, 
N.~Agrawal\,\orcidlink{0000-0003-0348-9836}\,$^{\rm 25}$, 
Z.~Ahammed\,\orcidlink{0000-0001-5241-7412}\,$^{\rm 133}$, 
S.~Ahmad\,\orcidlink{0000-0003-0497-5705}\,$^{\rm 15}$, 
S.U.~Ahn\,\orcidlink{0000-0001-8847-489X}\,$^{\rm 71}$, 
I.~Ahuja\,\orcidlink{0000-0002-4417-1392}\,$^{\rm 36}$
% ...

\section*{Affiliation Notes}

$^{\rm I}$ Deceased\\
$^{\rm II}$ Also at: Max-Planck-Institut fur Physik, Munich, Germany\\
$^{\rm III}$ Also at: Italian National Agency for New Technologies, Energy and Sustainable Economic Development (ENEA), Bologna, Italy\\
$^{\rm IV}$ Also at: Instituto de Fisica da Universidade de Sao Paulo\\
$^{\rm V}$ Also at: Dipartimento DET del Politecnico di Torino, Turin, Italy\\
$^{\rm VI}$ Also at: Department of Applied Physics, Aligarh Muslim University, Aligarh, India\\
$^{\rm VII}$ Also at: Institute of Theoretical Physics, University of Wroclaw, Poland\\
$^{\rm VIII}$ Also at: Facultad de Ciencias, Universidad Nacional Aut\'{o}noma de M\'{e}xico, Mexico City, Mexico\\

\section*{Collaboration Institutes}

$^{1}$ A.I. Alikhanyan National Science Laboratory (Yerevan Physics Institute) Foundation, Yerevan, Armenia\\
$^{2}$ AGH University of Krakow, Cracow, Poland\\
$^{3}$ Bogolyubov Institute for Theoretical Physics, National Academy of Sciences of Ukraine, Kiev, Ukraine\\
$^{4}$ Bose Institute, Department of Physics  and Centre for Astroparticle Physics and Space Science (CAPSS), Kolkata, India\\
$^{5}$ California Polytechnic State University, San Luis Obispo, California, United States\\
$^{6}$ Central China Normal University, Wuhan, China\\
$^{7}$ Centro de Aplicaciones Tecnol\'{o}gicas y Desarrollo Nuclear (CEADEN), Havana, Cuba\\
$^{8}$ Centro de Investigaci\'{o}n y de Estudios Avanzados (CINVESTAV), Mexico City and M\'{e}rida, Mexico\\
$^{9}$ Chicago State University, Chicago, Illinois, United States\\
$^{10}$ China Nuclear Data Center, China Institute of Atomic Energy, Beijing, China\\
$^{11}$ China University of Geosciences, Wuhan, China\\
$^{12}$ Chungbuk National University, Cheongju, Republic of Korea\\
$^{13}$ Comenius University Bratislava, Faculty of Mathematics, Physics and Informatics, Bratislava, Slovak Republic\\
$^{14}$ Creighton University, Omaha, Nebraska, United States\\
$^{15}$ Department of Physics, Aligarh Muslim University, Aligarh, India\\
$^{16}$ Department of Physics, Pusan National University, Pusan, Republic of Korea\\
$^{17}$ Department of Physics, Sejong University, Seoul, Republic of Korea\\
$^{18}$ Department of Physics, University of California, Berkeley, California, United States\\
$^{19}$ Department of Physics, University of Oslo, Oslo, Norway\\
$^{20}$ Department of Physics and Technology, University of Bergen, Bergen, Norway\\
$^{21}$ Dipartimento di Fisica, Universit\`{a} di Pavia, Pavia, Italy\\
$^{22}$ Dipartimento di Fisica dell'Universit\`{a} and Sezione INFN, Cagliari, Italy\\
$^{23}$ Dipartimento di Fisica dell'Universit\`{a} and Sezione INFN, Trieste, Italy\\
$^{24}$ Dipartimento di Fisica dell'Universit\`{a} and Sezione INFN, Turin, Italy\\
$^{25}$ Dipartimento di Fisica e Astronomia dell'Universit\`{a} and Sezione INFN, Bologna, Italy\\
$^{26}$ Dipartimento di Fisica e Astronomia dell'Universit\`{a} and Sezione INFN, Catania, Italy\\
$^{27}$ Dipartimento di Fisica e Astronomia dell'Universit\`{a} and Sezione INFN, Padova, Italy\\
$^{28}$ Dipartimento di Fisica `E.R.~Caianiello' dell'Universit\`{a} and Gruppo Collegato INFN, Salerno, Italy\\
$^{29}$ Dipartimento DISAT del Politecnico and Sezione INFN, Turin, Italy\\
$^{30}$ Dipartimento di Scienze MIFT, Universit\`{a} di Messina, Messina, Italy\\
$^{31}$ Dipartimento Interateneo di Fisica `M.~Merlin' and Sezione INFN, Bari, Italy\\
$^{32}$ European Organization for Nuclear Research (CERN), Geneva, Switzerland\\
$^{33}$ Faculty of Electrical Engineering, Mechanical Engineering and Naval Architecture, University of Split, Split, Croatia\\
$^{34}$ Faculty of Nuclear Sciences and Physical Engineering, Czech Technical University in Prague, Prague, Czech Republic\\
$^{35}$ Faculty of Physics, Sofia University, Sofia, Bulgaria\\
$^{36}$ Faculty of Science, P.J.~\v{S}af\'{a}rik University, Ko\v{s}ice, Slovak Republic\\
$^{37}$ Faculty of Technology, Environmental and Social Sciences, Bergen, Norway\\
$^{38}$ Frankfurt Institute for Advanced Studies, Johann Wolfgang Goethe-Universit\"{a}t Frankfurt, Frankfurt, Germany\\
$^{39}$ Fudan University, Shanghai, China\\
$^{40}$ Gangneung-Wonju National University, Gangneung, Republic of Korea\\
$^{41}$ Gauhati University, Department of Physics, Guwahati, India\\
$^{42}$ Helmholtz-Institut f\"{u}r Strahlen- und Kernphysik, Rheinische Friedrich-Wilhelms-Universit\"{a}t Bonn, Bonn, Germany\\
$^{43}$ Helsinki Institute of Physics (HIP), Helsinki, Finland\\
$^{44}$ High Energy Physics Group,  Universidad Aut\'{o}noma de Puebla, Puebla, Mexico\\
$^{45}$ Horia Hulubei National Institute of Physics and Nuclear Engineering, Bucharest, Romania\\
$^{46}$ HUN-REN Wigner Research Centre for Physics, Budapest, Hungary\\
$^{47}$ Indian Institute of Technology Bombay (IIT), Mumbai, India\\
$^{48}$ Indian Institute of Technology Indore, Indore, India\\
$^{49}$ INFN, Laboratori Nazionali di Frascati, Frascati, Italy\\
$^{50}$ INFN, Sezione di Bari, Bari, Italy\\
$^{51}$ INFN, Sezione di Bologna, Bologna, Italy\\
$^{52}$ INFN, Sezione di Cagliari, Cagliari, Italy\\
$^{53}$ INFN, Sezione di Catania, Catania, Italy\\
$^{54}$ INFN, Sezione di Padova, Padova, Italy\\
$^{55}$ INFN, Sezione di Pavia, Pavia, Italy\\
$^{56}$ INFN, Sezione di Torino, Turin, Italy\\
$^{57}$ INFN, Sezione di Trieste, Trieste, Italy\\
$^{58}$ Inha University, Incheon, Republic of Korea\\
$^{59}$ Institute for Gravitational and Subatomic Physics (GRASP), Utrecht University/Nikhef, Utrecht, Netherlands\\
$^{60}$ Institute of Experimental Physics, Slovak Academy of Sciences, Ko\v{s}ice, Slovak Republic\\
$^{61}$ Institute of Physics, Homi Bhabha National Institute, Bhubaneswar, India\\
$^{62}$ Institute of Physics of the Czech Academy of Sciences, Prague, Czech Republic\\
$^{63}$ Institute of Space Science (ISS), Bucharest, Romania\\
$^{64}$ Institut f\"{u}r Kernphysik, Johann Wolfgang Goethe-Universit\"{a}t Frankfurt, Frankfurt, Germany\\
$^{65}$ Instituto de Ciencias Nucleares, Universidad Nacional Aut\'{o}noma de M\'{e}xico, Mexico City, Mexico\\
$^{66}$ Instituto de F\'{i}sica, Universidade Federal do Rio Grande do Sul (UFRGS), Porto Alegre, Brazil\\
$^{67}$ Instituto de F\'{\i}sica, Universidad Nacional Aut\'{o}noma de M\'{e}xico, Mexico City, Mexico\\
$^{68}$ iThemba LABS, National Research Foundation, Somerset West, South Africa\\
$^{69}$ Jeonbuk National University, Jeonju, Republic of Korea\\
$^{70}$ Johann-Wolfgang-Goethe Universit\"{a}t Frankfurt Institut f\"{u}r Informatik, Fachbereich Informatik und Mathematik, Frankfurt, Germany\\
$^{71}$ Korea Institute of Science and Technology Information, Daejeon, Republic of Korea\\
$^{72}$ Laboratoire de Physique Subatomique et de Cosmologie, Universit\'{e} Grenoble-Alpes, CNRS-IN2P3, Grenoble, France\\
$^{73}$ Lawrence Berkeley National Laboratory, Berkeley, California, United States\\
$^{74}$ Lund University Department of Physics, Division of Particle Physics, Lund, Sweden\\
$^{75}$ Nagasaki Institute of Applied Science, Nagasaki, Japan\\
$^{76}$ Nara Women{'}s University (NWU), Nara, Japan\\
$^{77}$ National and Kapodistrian University of Athens, School of Science, Department of Physics , Athens, Greece\\
$^{78}$ National Centre for Nuclear Research, Warsaw, Poland\\
$^{79}$ National Institute of Science Education and Research, Homi Bhabha National Institute, Jatni, India\\
$^{80}$ National Nuclear Research Center, Baku, Azerbaijan\\
$^{81}$ National Research and Innovation Agency - BRIN, Jakarta, Indonesia\\
$^{82}$ Niels Bohr Institute, University of Copenhagen, Copenhagen, Denmark\\
$^{83}$ Nikhef, National institute for subatomic physics, Amsterdam, Netherlands\\
$^{84}$ Nuclear Physics Group, STFC Daresbury Laboratory, Daresbury, United Kingdom\\
$^{85}$ Nuclear Physics Institute of the Czech Academy of Sciences, Husinec-\v{R}e\v{z}, Czech Republic\\
$^{86}$ Oak Ridge National Laboratory, Oak Ridge, Tennessee, United States\\
$^{87}$ Ohio State University, Columbus, Ohio, United States\\
$^{88}$ Physics department, Faculty of science, University of Zagreb, Zagreb, Croatia\\
$^{89}$ Physics Department, Panjab University, Chandigarh, India\\
$^{90}$ Physics Department, University of Jammu, Jammu, India\\
$^{91}$ Physics Program and International Institute for Sustainability with Knotted Chiral Meta Matter (WPI-SKCM$^{2}$), Hiroshima University, Hiroshima, Japan\\
$^{92}$ Physikalisches Institut, Eberhard-Karls-Universit\"{a}t T\"{u}bingen, T\"{u}bingen, Germany\\
$^{93}$ Physikalisches Institut, Ruprecht-Karls-Universit\"{a}t Heidelberg, Heidelberg, Germany\\
$^{94}$ Physik Department, Technische Universit\"{a}t M\"{u}nchen, Munich, Germany\\
$^{95}$ Politecnico di Bari and Sezione INFN, Bari, Italy\\
$^{96}$ Research Division and ExtreMe Matter Institute EMMI, GSI Helmholtzzentrum f\"ur Schwerionenforschung GmbH, Darmstadt, Germany\\
$^{97}$ Saga University, Saga, Japan\\
$^{98}$ Saha Institute of Nuclear Physics, Homi Bhabha National Institute, Kolkata, India\\
$^{99}$ School of Physics and Astronomy, University of Birmingham, Birmingham, United Kingdom\\
$^{100}$ Secci\'{o}n F\'{\i}sica, Departamento de Ciencias, Pontificia Universidad Cat\'{o}lica del Per\'{u}, Lima, Peru\\
$^{101}$ Stefan Meyer Institut f\"{u}r Subatomare Physik (SMI), Vienna, Austria\\
$^{102}$ SUBATECH, IMT Atlantique, Nantes Universit\'{e}, CNRS-IN2P3, Nantes, France\\
$^{103}$ Sungkyunkwan University, Suwon City, Republic of Korea\\
$^{104}$ Suranaree University of Technology, Nakhon Ratchasima, Thailand\\
$^{105}$ Technical University of Ko\v{s}ice, Ko\v{s}ice, Slovak Republic\\
$^{106}$ The Henryk Niewodniczanski Institute of Nuclear Physics, Polish Academy of Sciences, Cracow, Poland\\
$^{107}$ The University of Texas at Austin, Austin, Texas, United States\\
$^{108}$ Universidad Aut\'{o}noma de Sinaloa, Culiac\'{a}n, Mexico\\
$^{109}$ Universidade de S\~{a}o Paulo (USP), S\~{a}o Paulo, Brazil\\
$^{110}$ Universidade Estadual de Campinas (UNICAMP), Campinas, Brazil\\
$^{111}$ Universidade Federal do ABC, Santo Andre, Brazil\\
$^{112}$ Universitatea Nationala de Stiinta si Tehnologie Politehnica Bucuresti, Bucharest, Romania\\
$^{113}$ University of Derby, Derby, United Kingdom\\
$^{114}$ University of Houston, Houston, Texas, United States\\
$^{115}$ University of Jyv\"{a}skyl\"{a}, Jyv\"{a}skyl\"{a}, Finland\\
$^{116}$ University of Kansas, Lawrence, Kansas, United States\\
$^{117}$ University of Liverpool, Liverpool, United Kingdom\\
$^{118}$ University of Science and Technology of China, Hefei, China\\
$^{119}$ University of South-Eastern Norway, Kongsberg, Norway\\
$^{120}$ University of Tennessee, Knoxville, Tennessee, United States\\
$^{121}$ University of the Witwatersrand, Johannesburg, South Africa\\
$^{122}$ University of Tokyo, Tokyo, Japan\\
$^{123}$ University of Tsukuba, Tsukuba, Japan\\
$^{124}$ Universit\"{a}t M\"{u}nster, Institut f\"{u}r Kernphysik, M\"{u}nster, Germany\\
$^{125}$ Universit\'{e} Clermont Auvergne, CNRS/IN2P3, LPC, Clermont-Ferrand, France\\
$^{126}$ Universit\'{e} de Lyon, CNRS/IN2P3, Institut de Physique des 2 Infinis de Lyon, Lyon, France\\
$^{127}$ Universit\'{e} de Strasbourg, CNRS, IPHC UMR 7178, F-67000 Strasbourg, France, Strasbourg, France\\
$^{128}$ Universit\'{e} Paris-Saclay, Centre d'Etudes de Saclay (CEA), IRFU, D\'{e}partment de Physique Nucl\'{e}aire (DPhN), Saclay, France\\
$^{129}$ Universit\'{e}  Paris-Saclay, CNRS/IN2P3, IJCLab, Orsay, France\\
$^{130}$ Universit\`{a} degli Studi di Foggia, Foggia, Italy\\
$^{131}$ Universit\`{a} del Piemonte Orientale, Vercelli, Italy\\
$^{132}$ Universit\`{a} di Brescia, Brescia, Italy\\
$^{133}$ Variable Energy Cyclotron Centre, Homi Bhabha National Institute, Kolkata, India\\
$^{134}$ Warsaw University of Technology, Warsaw, Poland\\
$^{135}$ Wayne State University, Detroit, Michigan, United States\\
$^{136}$ Yale University, New Haven, Connecticut, United States\\
$^{137}$ Yildiz Technical University, Istanbul, Turkey\\
$^{138}$ Yonsei University, Seoul, Republic of Korea\\
$^{139}$ Affiliated with an institute formerly covered by a cooperation agreement with CERN\\
$^{140}$ Affiliated with an international laboratory covered by a cooperation agreement with CERN.\\

\end{flushleft} 
  
\end{document}